\documentclass[conference, 9pt]{IEEEtran}
\IEEEoverridecommandlockouts

\usepackage{cite}
\usepackage{amsmath,amssymb,amsfonts}
\usepackage{algorithmic}
\usepackage{graphicx}
\usepackage{multirow}
\usepackage{booktabs}
\usepackage{soul}
\usepackage{textcomp, hyperref}
\usepackage{xcolor}
\def\BibTeX{{\rm B\kern-.05em{\sc i\kern-.025em b}\kern-.08em
    T\kern-.1667em\lower.7ex\hbox{E}\kern-.125emX}}
\begin{document}



\title{Editing Music with Melody and Text:\\ Using ControlNet for  Diffusion Transformer}

\author{

\IEEEauthorblockN{
    Siyuan Hou$^{1,2}$, 
    Shansong Liu$^{2}$, 
    Ruibin Yuan$^{3}$, 
    Wei Xue$^{3}$, 
    Ying Shan$^{2}$,
    Mangsuo Zhao$^{1}$, 
    Chao Zhang$^{1,*}$
    \thanks{$^{*}$Corresponding author. This work was done when Siyuan Hou was an intern at Tencent ARC Lab.} 
}

\IEEEauthorblockA{
    \textit{$^{1}$ Tsinghua University}, \textit{$^{2}$ ARC Lab, Tencent PCG}, \textit{$^{3}$ Hong Kong University of Science and Technology} \\
}

\textit{
    housy24@mails.tsinghua.edu.cn;
    shansongliu@tencent.com;
    cz277@tsinghua.edu.cn}
}

\maketitle

\begin{abstract}
Despite the significant progress in controllable music generation and editing, challenges remain in the quality and length of generated music due to the use of Mel-spectrogram representations and UNet-based model structures. To address these limitations, we propose a novel approach using a Diffusion Transformer (DiT) augmented with an additional control branch using ControlNet. This allows for long-form and variable-length music generation and editing controlled by text and melody prompts.
For more precise and fine-grained melody control, we introduce a novel top-$k$ constant-Q Transform representation as the melody prompt, reducing ambiguity compared to previous representations (e.g., chroma), particularly for music with multiple tracks or a wide range of pitch values. To effectively balance the control signals from text and melody prompts, we adopt a curriculum learning strategy that progressively masks the melody prompt, resulting in a more stable training process.
Experiments have been performed on text-to-music generation and music-style transfer tasks using open-source instrumental recording data. The results demonstrate that by extending StableAudio, a pre-trained text-controlled DiT model, our approach enables superior melody-controlled editing while retaining good text-to-music generation performance. These results outperform a strong MusicGen baseline in terms of both text-based generation and melody preservation for editing. Audio examples can be found at \href{https://stable-audio-control.github.io}{https://stable-audio-control.github.io}.

\end{abstract}



\section{Introduction}
With the rapid development of deep generative models, significant progress has been made in text-guided generation tasks across multiple modalities, including text-to-image, text-to-audio, and text-to-video. In the realm of text-to-music generation, several models, such as MusicLM \cite{musiclm}, MusicGen \cite{musicgen}, JASCO \cite{jasco}, and JEN-1 \cite{jen-1}, have shown to produce high-quality music based on textual prompts, which typically describe a more abstract and high-level form of global attributes, such as style, theme, and genre \cite{musiccontrolnet}. Additionally, some models have explored more flexible conditioning mechanisms in music generation tasks, such as conditioning on musical audio contexts\cite{diff-a-riff, stemgen, multi-source, singsong}. Recently, efforts in music editing have emerged, including Coco-Mulla \cite{Coco-Mulla} and DITTO \cite{ditto}, that use temporally aligned music prompts to achieve greater flexibility in controlling the finer attributes of the generated music.

Existing music generation models face two limitations. First, the use of UNet-based Diffusion models \cite{musicldm, musicti, auffusion} to generate music via Mel-spectrogram representations imposes significant constraints on both the length and quality of the generated music. The fixed-length output of spectrograms, required to accommodate the 2D convolution structure and downsampling factors, hinders the generation of precise, variable-length audio. Additionally, converting Mel-spectrograms to waveforms via a separately trained vocoder often introduces errors during the conversion process, which further degrades the quality of the final audio output.
Second, many existing methods use melody prompts that fail to include complete melody information, leading to poor melody retention. For instance, Music ControlNet \cite{musiccontrolnet} employs a one-hot 12-pitch-class \textit{chromagram} as the melody prompt, which cannot accurately capture pitch changes across multiple octaves and struggles to represent the melody of multi-track music effectively.

To address the aforementioned challenges, we propose an improved controllable music editing model designed to generate music based on both text and melody prompts. Inspired by the StableAudio models \cite{stable-audio,stable-audio-2}, we adopt the Diffusion Transformer (DiT) model structure with a timing conditioning method as the primary framework, enabling the generation of variable-length music. For precise melody control, a control branch based on ControlNet is introduced, following the approach used in Music ControlNet \cite{musiccontrolnet}, to adapt the Diffusion component of the DiT model.
In contrast to the chroma features commonly used in previous studies, we propose to use a top-$k$ constant-Q Transform (CQT) feature to provide more flexible and accurate melody representations. This feature helps disambiguate and render music with multiple tracks or a broader range of pitch values. Furthermore, drawing inspiration from the JEN-1 composer \cite{jen-composer}, a progressive curriculum masking strategy is implemented to improve the learning of melody prompts, achieving a more effective balance between text and melody prompts during training.
As a result, our model not only can perform prompt based music style transfer but also render and generate music content more precisely using melody prompts. Experimental results show that our model delivers competitive performance across various subjective and objective metrics, showcasing enhanced music editing abilities without considerable loss in text-to-music generation quality.

Our main contributions are as follows:
\begin{itemize}
\item To the best of our knowledge, we are the first to apply ControlNet \cite{controlnet} to the DiT model for music tasks, enabling precise, time-varying control over melody prompts.
\item We introduce top-$k$ CQT, a flexible and effective representation of melody information based on constant-Q Transform features, which retains more detailed melody information and enhances the model’s ability to preserve melodies.
\item We propose a progressive curriculum masking strategy to stabilize training and improve music editing performance.
\end{itemize}


\section{Background}
\subsection{Diffusion Model}
The diffusion model \cite{ddpm} is a type of generative approach based on probabilistic models, where the core idea is to decompose the generation process into a series of incremental random diffusion steps. By learning to reverse this process of adding random noise, the diffusion model gradually acquires the ability to generate target data from noise. Due to its high-quality generation and stable process, the diffusion model architecture has been adopted by various audio generation models, including Moûsai \cite{mousai}, AudioLDM \cite{audioldm, audioldm2}, MusicLDM \cite{musicldm, mt-musicldm}, and MusicMagus \cite{musicmagus}.


In the forward process, a stochastic process defined by a Markov chain increasingly adds noise to the clean data $\mathbf{x}_0$:
\begin{equation}
\begin{aligned}
&q(\mathbf x_{1:T}|\mathbf x_0)=\prod\nolimits_{t=1}^Tq(\mathbf x_{t}|\mathbf x_{t-1}) \\
q(\mathbf x_{t}|&\mathbf x_{t-1})=\mathcal N(\mathbf x_t;\sqrt{1-\beta_t}\mathbf x_{t-1},\beta_t\mathbf I),
\end{aligned}
\end{equation}
where $\beta_1,\cdots,\beta_T$ are a sequence of predetermined noise schedule that define the noise level within the forward diffusion process.

During training, diffusion models learn the reverse process with learnable parameters $\theta$, which iteratively denoise a random sample $\mathbf{x}_T\sim\mathcal N(\mathbf{0},\mathbf{I})$ to a data sample $\mathbf{x}_0\sim p(\mathbf{x}_0)$:
\begin{equation}
\begin{aligned}
p_{\theta}(&\mathbf x_{0:T-1}|\mathbf x_T)=\prod\nolimits_{t=1}^T p_\theta(\mathbf x_{t-1}|\mathbf x_{t}) \\
p_\theta(\mathbf x_{t-1}&|\mathbf x_{t})=\mathcal N(\mathbf x_{t-1};\mu_\theta(\mathbf x_t,t), \Sigma_\theta(\mathbf x_t,t)).
\end{aligned}
\end{equation}
This denoising process can be trained by minimizing the mean squared error of the predicted noise $\epsilon\sim\mathcal N(\mathbf{0}, \mathbf{I})$. Recently, the v-objective diffusion has been proposed \cite{v-objective}, which estimates the model $\hat{\mathbf{v}}_{t}=f_\theta(\mathbf{x}_{t}, t)$ by minimizing the following objective:
\begin{equation}
    \mathbb{E} _{t\sim[0,1],\mathbf x_t}
    \ \|f_\theta(\alpha_t \mathbf x_0 + \sigma_t\epsilon,t)-\mathbf v_{\sigma_t}\|_2^2,
\end{equation} where $\mathbf v_t={d \mathbf x_t}/{d t}=\alpha_t\epsilon-\sigma_t\mathbf x_0$ and $t\sim[0,1]$ is noise schedule, $\alpha_t=\cos(0.5\pi t)$ and $\sigma_t=\sin(0.5\pi t)$.

\subsection{Diffusion Transformer (DiT)}
DiT \cite{dit} is a generative model that integrates the Diffusion model with Transformer architecture \cite{transformer}. Unlike traditional diffusion models that rely on the UNet architecture \cite{ldm}, DiT uses a Transformer encoder as the denoising network. By leveraging the self-attention mechanism of the Transformer, which excels at capturing long-range dependencies in data, it is particularly effective for modelling the complex structure of music, allowing it to generate coherent and consistent compositions. StableAudio 2 \cite{stable-audio-2} applies DiT to audio generation, enabling high-quality, long-form music generation.


\begin{figure*}[t]
\centering
\begin{minipage}[b]{0.42\linewidth}
  \centering  
  \centerline{\vspace{1.5cm}\hspace{-1.5cm}\includegraphics[width=1.1\linewidth]{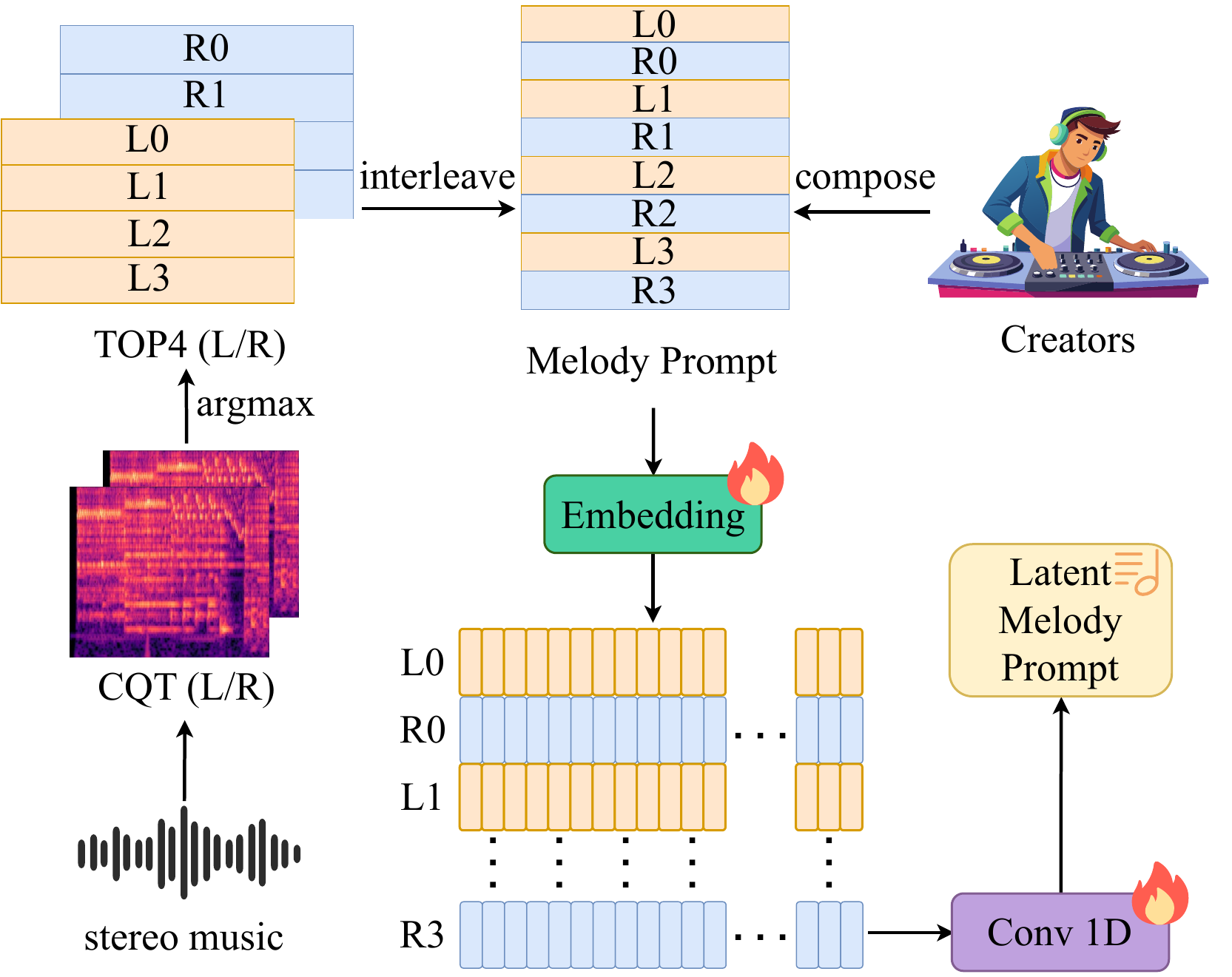}}
  \centerline{\hspace{-1.5cm} \footnotesize{(a) Processing pipeline for melody prompts.}}\medskip
\end{minipage}
\hspace{1.0cm}
\begin{minipage}[b]{0.4\linewidth}
  \centering  
  \centerline{\includegraphics[width=1.2\linewidth]{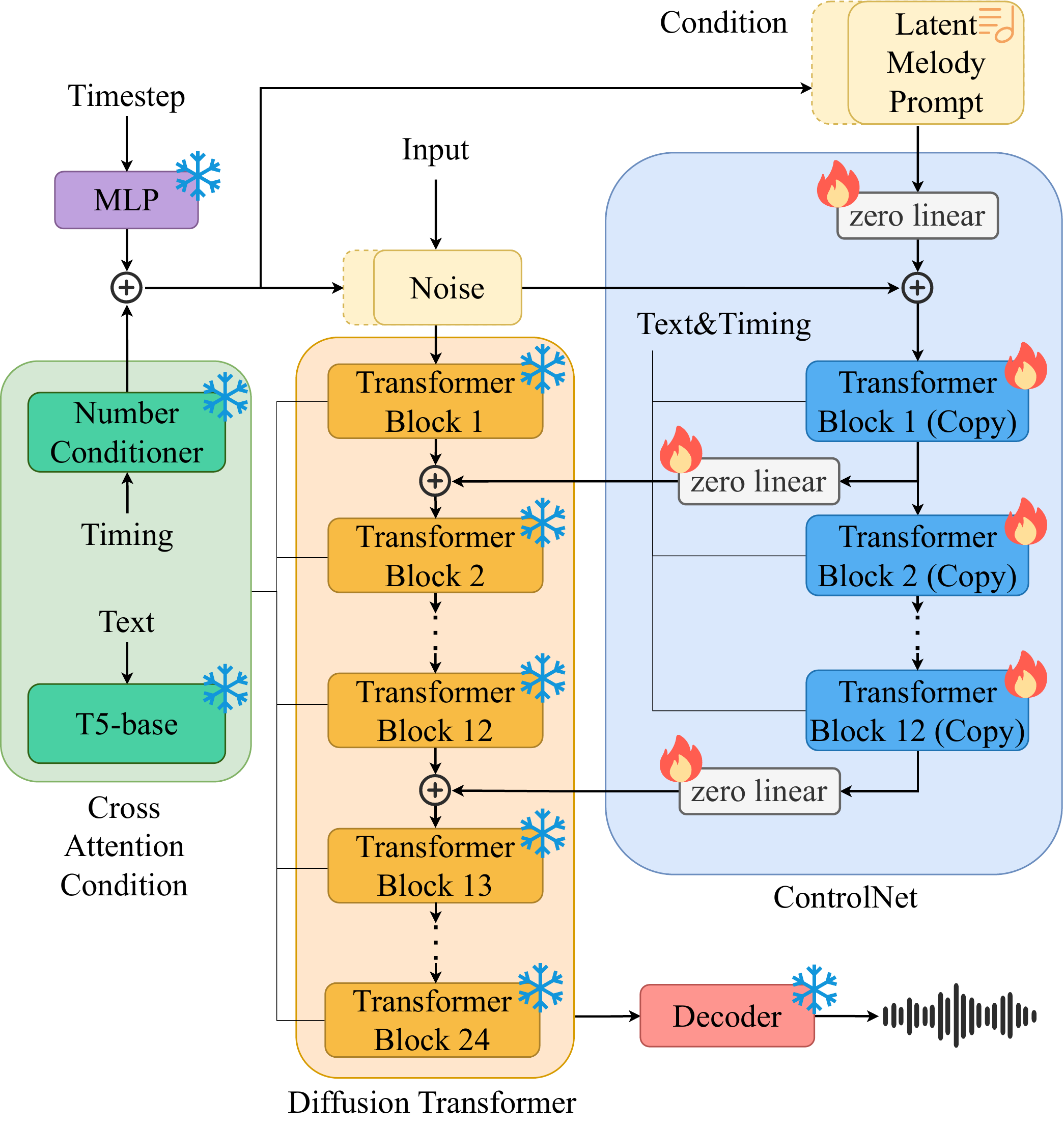}}
  \vspace{0.3cm}
  \centerline{\footnotesize{(b) Our model for music editing.}}\medskip
\end{minipage}
\caption{Overview of our Music Editing Model: (a) Melody prompt processing pipeline. The top-$k$ CQT of the stereo audio is computed, and the prominent components are selected to form the melody prompt, which can also be manually composed. The latent melody prompt is then derived through further processing using an embedding layer and 1D convolutions. (b) The architecture of our model. The model primarily consists of DiT and ControlNet. Various conditioning inputs are supplied via two methods: prepending and cross-attention. During training, only the ControlNet and the structure responsible for extracting melody prompts are fine-tuned. In the figure, \includegraphics[height=1.5\fontcharht\font`X]{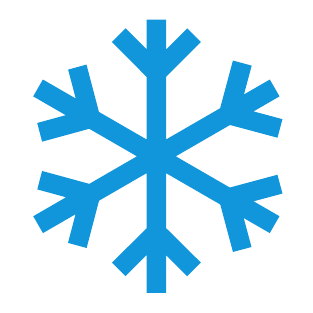} indicates frozen components, while \includegraphics[height=1.5\fontcharht\font`X]{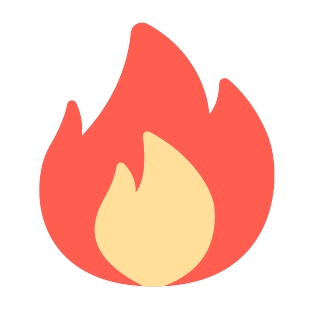} indicates components that are finetuned.}
\label{fig1}
\end{figure*}

\section{Proposed Method}
\subsection{Diffusion Transformer with ControlNet}
In order to model long-form and variable-length music, we adopt StableAudio Open \cite{stable-audio-open} as the backbone structure, which uses the DiT model. By combining DiT's sequence modelling abilities and the low-latency advantages of an audio codec, this framework can generate high-quality stereo audio up to 47 seconds in length, guided by text prompts. In addition, thanks to StableAudio’s timing conditioning mechanism \cite{stable-audio}, the model can generate music with variable lengths, offering greater flexibility in music generation.


Building on this foundation and inspired by Music ControlNet \cite{musiccontrolnet}, we explore adding a control branch to diffusion models for time-varying melody control. Music ControlNet enables control over musical attributes such as melody and dynamics by adapting ControlNet to the music domain. However, since Music ControlNet uses a UNet-structured diffusion model and ControlNet was designed for the UNet architecture, the original ControlNet is not directly compatible with the DiT structure we employed. For instance, DiT lacks the explicit symmetrical ``encoder-decoder'' structure present in UNet, making components like skip connections, which are utilized by the original ControlNet, unsuitable in this context.


To address this issue, we draw inspiration from the ControlNet-Transformer structure used in Pixart-$\delta$ \cite{pixart} and adapt it to the DiT architecture of StableAudio (as shown in Fig. \ref{fig1}). Specifically, ControlNet-Transformer replicates the first $N$ Transformer blocks as the control branch. The output of the $i^\text{th}$ trainable copy block, after passing through an additional zero-initialized linear layer, is combined with the output of the $i^\text{th}$ frozen block and used as the input to the $(i+1)^\text{th}$ frozen block. This design maintains the original connection structure of the DiT model, enabling seamless integration of the ControlNet structure while preserving the core advantages of the Transformer structure.

\subsection{Top-\texorpdfstring{$k$}{k} CQT Representation for Melody Information}
Finding a precise and flexible representation of melody information is critical to achieving time-varying control over music melody. Previous studies, such as Music ControlNet \cite{musiccontrolnet} and DITTO \cite{ditto, ditto2}, have used a one-hot 12-pitch-class chromagram as the melody control condition, which has two primary disadvantages. First, it only captures pitch changes within a single octave, meaning the model struggles to preserve melodies involving pitch changes across multiple octaves and cannot accurately infer absolute pitch information. Second, the argmax operation used in this method records only the most prominent pitch class for each frame, which limits its ability to extract complete melody information in music with multiple interwoven tracks. As a result, the generated melody may deviate significantly from the original, capturing only the overall melody envelope rather than the full harmonic details.

To address the issues mentioned, we propose a more precise and flexible melody representation (Fig. \ref{fig1}) by extending CQT \cite{cqt} to have the top-$k$ most prominent pitch values. Since CQT’s centre frequencies are distributed exponentially, this representation aligns well with the frequency distribution of musical scales, making it highly effective for processing musical signals \cite{cqt-2017, cqt-2024}.
For stereo audio input, we first compute the CQT with 128 bins for both left and right channels, then apply an argmax operation to retain the 4 most prominent pitches per frame in each channel. This results in 8 prominent pitches, interleaved and stacked, forming the melody prompt $\mathbf{c} \in \mathbb{R}^{8 \times Tf_k}$, where $T$ is the length of the audio in seconds and $f_k$ is the frame rate after CQT computation, with each pixel $\mathbf{c}_{i,j} \in {1,2,\cdots,128}$. As in Music ControlNet, we apply a biquadratic high-pass filter with a cut-off at Middle C (261.2 Hz) before computing the CQT.
Since top-$k$ CQT retains absolute pitch information and captures the melody details of multiple tracks, it significantly improves the precision of melody control. Meanwhile, top-$k$ CQT maintains the flexibility of chroma representation, as it uses human-understandable pitch indices as melody prompts. This allows for not only extracting melody from audio but also converting music scores into melody control data or directly creating melodic representations as inputs for the model.


To further enhance the extraction and representation of melody information, we use the melody prompt $\mathbf{c}$ as an index, transforming it from a representation of prominent pitch classes into a higher-dimensional latent melody prompt through pitch-specific trainable embeddings. Next, a series of 1D convolution layers are applied to downsample the melody information, aligning it with the input shape of ControlNet, thus enabling precise control over the melody in the generated music.

\subsection{Progressive Curriculum Masking Strategy}
\label{mask}
While a precise melody prompt allows for greater control over the music's melody, it also introduces challenges to the model's creative flexibility. On one hand, overly precise melody prompts may incorporate additional musical elements, such as timbre, into the conditional information. In this case, the melody representation functions more like a compressed version of the target audio rather than a pure melody representation. On the other hand, since the melody prompt is temporally aligned with the target audio, the model is more likely to learn the direct relationship between the melody prompt and the audio, rather than relying on the high-level text prompt. As a result of these two factors, the model may become adept at reconstructing the target audio from precise melody prompts, but less capable of generating music guided by both melody and text prompts.


To address this issue, adjustments need to be made to the melody prompts during training. Inspired by JEN-1 Composer\cite{jen-composer}, we introduce a progressive curriculum masking strategy for the melody prompts to gradually enhance the model's ability to utilize them for music modeling. The masking is applied in two directions: frame-wise and pitch-class-wise. Initially, all melody prompts are masked, allowing the model to learn music generation with empty melody prompts. As training progresses, the mask ratio in the frame direction is gradually reduced, enabling the model to progressively learn to incorporate melody prompts.
It is important to note that the mask ratio does not decrease monotonically during training, as the probability of sampling smaller mask ratios increases over time, while a small probability of sampling larger mask ratios remains. After the initial full-mask phase, the top-1 melody prompt is retained, while the top-2, top-3, and top-4 melody prompts are randomly masked and shuffled. This curriculum masking strategy helps the model learn to use melody prompts more robustly, preventing overfitting to simple reconstruction tasks and avoiding the loss of its text-to-music generation ability. Ultimately, this approach strikes a balance between controlling generation with text and melody prompts, ensuring the model retains its text-to-music generation capabilities while learning music editing tasks.

\begin{table*}[t!]
\caption{The objective results on the Song Describer Dataset. For the Text-to-Music task, music is generated based on a text prompt. For the Music Editing task, music is generated using both a text prompt and a melody prompt extracted from a given audio sample. The \textbf{top-1 performance} is highlighted in bold, and the \underline{top-2 performance} is underlined.}
\setlength{\tabcolsep}{5pt}
\centering 

\begin{tabular}{c|ccc|cccc}
\toprule
\multirow{2}{*}{Model} & \multicolumn{3}{c|}{\textbf{Text-to-Music}} & \multicolumn{4}{c}{\textbf{Music Editing}} \\
 & FD$_\text{openl3}\downarrow$ & KL$_\text{passet}\downarrow$ & CLAP$_\text{score}\uparrow$ & Melody acc $\uparrow$ & FD$_\text{openl3}\downarrow$ & KL$_\text{passet}\downarrow$ & CLAP$_\text{score} \uparrow$ \\
\midrule
MusicGen-stereo-melody & 218.82 & 0.811 & 0.249 &  42.3\%  &199.35 & 0.432 & 0.348 \\
MusicGen-stereo-melody-large & 201.53 & 0.707 & 0.279 & 44.7\% & 190.79 & 0.384 & 0.354\\
\midrule
Ours w/ cross attention & \textbf{116.71} & \textbf{0.531} & \textbf{0.417} & 9.37\% & 115.10 & 0.587 & \textbf{0.412} \\
Ours w/o masking strategy & — & — & — & \ul{54.9\%} & \textbf{86.060} & \textbf{0.232} & 0.393 \\
Ours & \ul{158.53} & \ul{0.623} & \ul{0.375} & \textbf{56.6\%} & \ul{97.734} & \ul{0.265} & \ul{0.396}\\
\bottomrule
\end{tabular}

\label{table1}
\end{table*}

\section{Experimental Setup}
\subsection{Data}
Our training data are sourced from four public music datasets: MTG\cite{mtg}, FMA\cite{fma}, MTT\cite{mtt}, and Wikimute\cite{wikimute}. Since our focus is on generating and editing instrumental music (\textit{i.e.}, no singing), we use the PANNs \cite{panns} tagger to classify the tracks and filter out the portions with vocals. Furthermore, as these datasets primarily contain music tags rather than free-form descriptions, we leverage SALMONN-13B \cite{salmonn} to generate multiple captions for each filtered audio sample. After processing, we compile a dataset consisting of 59,955 recordings, totaling 2,239.7 hours of high-quality text-music paired data. For evaluation, we select the no-singing subset from the Song Describer \cite{songdescriber} dataset, which is the same benchmark used in the stable audio evaluations \cite{stable-audio,stable-audio-2,stable-audio-open}.

\subsection{Model}
For our pre-trained diffusion transformer model, we utilized the checkpoint from StableAudio Open \cite{stable-audio-open}, which includes both the DiT model and an autoencoder. ControlNet clones half of the pre-trained DiT, creating a control branch with a total of 12 Transformer blocks. All input audio is resampled and processed at 44.1 kHz in stereo format. Melody prompts are extracted using the CQT with a hop length of 512 and an fmin corresponding to a MIDI note number of 0 (8.18 Hz). Each octave contains 12 bins, resulting in a total of 128 bins, which perfectly aligns with the 128 MIDI notes.

During training, we utilize the v-objective \cite{v-objective} method, enabling the model to predict noise increments from noisy ground-truth data. The AdamW optimizer is used with a learning rate of 5e-5, along with an InverseLR scheduler set to a power of 0.5. The DiT parameters remain frozen, while only the ControlNet and the layers responsible for converting latent melody prompt are fine-tuned. Training is guided by mean squared error loss and continues until convergence, which takes approximately 3 days on 4 V100 GPUs, with a mini-batch size of 8 per GPU. For inference, we employ 250-step DPM-Solver++ \cite{dpmsolver++} sampling, using classifier-free guidance \cite{cfg} with a scale of 7, applied exclusively to global text prompt control.

\subsection{Evaluation}

To assess time-varying melody controllability, adherence to global text control, and overall audio realism, we use a combination of subjective and objective metrics.
For the objective metrics, melody accuracy evaluates the frame-wise alignment between pitch classes of the input melody control and those extracted from the generated audio, reflecting the model's ability to preserve the melody during music editing. We use $\text{FD}_\text{openl3}$ and $\text{KL}_\text{passt}$ to evaluate the difference between the model-generated audio and the ground-truth reference. Additionally, $\text{CLAP}_\text{score}$ is used to assess the consistency between the generated audio and the provided text prompt.
For the subjective metrics, we employ mean opinion scores (MOS) on a 5-point scale to evaluate both the alignment of the generated music with the text prompt and the overall audio quality. In total, 20 volunteers from different age groups participated in the evaluation. Among them, one-third have had prior experience learning musical instruments or received musical education.

Our experiments evaluate two tasks: text-to-music generation and music editing. For the music editing task, we generate music based on two conditions: a music prompt extracted from the target audio and the corresponding text prompt. We select MusicGen-stereo-melody as baseline, as it is one of the most competitive open-source models for stereo music generation and editing. Additionally, we conduct an ablation study to explore various architectures for handling music prompts and different masking strategies. Specifically, we examine alternative ControlNet structures by replacing information injection via addition with cross-attention mechanisms.


\section{Experimental Results}
\subsection{Objective Results}
The objective results in Table~\ref{table1} demonstrate that our model successfully balances text-to-music generation and music editing, delivering strong performance in both tasks. Although our model degrades text-to-music performance to improve music editing performance compared to StableAudio Open, it still significantly outperforms MusicGen. In the music editing task, our model excels across all four evaluation metrics, highlighting its proficiency in music editing while retaining a solid level of text-to-music generation capability.

\begin{table}[ht]
\caption{The subjective results on the Song Describer dataset. TF and OVL represent text fit and overall quality, respectively.} 
\centering 
\begin{tabular}{c|cc|cc}
\toprule
\multirow{2}{*}{Model} & \multicolumn{2}{c|}{\textbf{Text-to-Music}} & \multicolumn{2}{c}{\textbf{Music Editing}} \\
 & TF & OVL & TF & OVL \\
\midrule
MusicGen-stereo-melody & 2.8 & 2.5 & 3.2  & 3.3\\
MusicGen-stereo-melody-large  & 3.3 & \textbf{3.6} & 3.5 & 3.5\\
Ours & \textbf{3.4} & \textbf{3.6} & \textbf{3.7} & \textbf{3.8} \\
\bottomrule
\end{tabular}
\label{table2}
\end{table}

\subsection{Subjective Results}
Table \ref{table2} presents the average MOS scores for text-music consistency and audio quality. The subjective results clearly show that our model outperforms MusicGEN in both tasks, with a particularly notable advantage in the music editing task.

\subsection{Ablation Study}
The results of the ablation study indicate that although injecting prompts via cross-attention aligns better with the Transformer structure and improves the quality of the generated music, it nearly eliminates the model’s music editing ability and struggles to balance the two types of prompts, essentially functioning as a pure text-to-music model, as reflected by the lower melody accuracy. 

Regarding the masking strategy, without it, the model cannot effectively maintain both text-to-music generation and music editing capabilities.  Without the masking strategy, the model is consistently trained on a complete melody prompt, which hinders its ability to perform text-to-music tasks that rely solely on a text prompt without a melody prompt. Although the model performs reasonably well on music editing metrics, this does not imply that it possesses strong music editing capabilities. As previously mentioned, during training, without the masking strategy, the model behaves more like it is reconstructing the target audio based on the melody prompt, rather than performing true music editing. 

Therefore, even though these two models in the ablation study exhibit top performance on many metrics, they fail to properly balance the control effects of the text prompt and the melody prompt.

\section{Conclusion}
In this paper, we propose a novel music generation and editing approach designed to support long-form and variable-length music as well as both text and melody prompts. To achieve this, we integrated ControlNet into the DiT structure and introduced a novel top-$k$ CQT representation as the melody prompt, providing more precise and fine-grained melody control. Both subjective and objective results show that our model, while maintaining strong text-to-music generation abilities, excels in music editing and achieves a more balanced control between text and melody prompts.

\bibliographystyle{IEEEtran}
\bibliography{reference}

\end{document}